\documentclass[letterpaper,11pt]{article}

\usepackage{graphicx,natbib,here,setspace,multirow}
\usepackage{amsmath,amssymb,amsfonts}
\usepackage{charter}

\usepackage[letterpaper, left=1.1in, top=1.1in, right=1.1in, bottom=1in, nohead,includefoot, verbose, ignoremp]{geometry}

\usepackage[compact]{titlesec}
\titlespacing{\section}{0pt}{5pt}{3pt}
\titlespacing{\subsection}{0pt}{5pt}{3pt}
\titlespacing{\subsubsection}{0pt}{4pt}{3pt}
\bibpunct{(}{)}{;}{a}{}{,}

\def\y{\mbox{\boldmath$y$}}
\def\w{\mbox{\boldmath$w$}}
\def\e{\mbox{\boldmath$e$}}
\def\a{\mbox{\boldmath$a$}}
\def\g{\mbox{\boldmath$g$}}
\def\h{\mbox{\boldmath$h$}}
\def\q{\mbox{\boldmath$q$}}
\def\x{\mbox{\boldmath$x$}}
\def\w{\mbox{\boldmath$w$}}
\def\A{\mbox{\boldmath$A$}}
\def\I{\mbox{\boldmath$I$}}
\def\V{\mbox{\boldmath$V$}}
\def\O{\mbox{\boldmath$O$}}
\def\Q{\mbox{\boldmath$Q$}}
\def\K{\mbox{\boldmath$K$}}

\def\S{\mbox{\boldmath$S$}}
\def\D{\mbox{\boldmath$D$}}
\def\bmu{\mbox{\boldmath$\mu$}}
\def\bfeta{\mbox{\boldmath$\eta$}}
\def\bbeta{\mbox{\boldmath$\beta$}}
\def\beps{\mbox{\boldmath$\varepsilon$}}
\def\btheta{\mbox{\boldmath$\theta$}}
\def\bxi{\mbox{\boldmath$\xi$}}
\def\bomega{\mbox{\boldmath$\omega$}}
\def\bPhi{\mbox{\boldmath$\Phi$}}
\def\bSigma{\mbox{\boldmath$\Sigma$}}
\def\bLambda{\mbox{\boldmath$\Lambda$}}
\def\bOmega{\mbox{\boldmath$\Omega$}}
\def\zero{\mbox{\boldmath$0$}}
\def\one{\mbox{\boldmath$1$}}
\def\diag{\mathrm{diag}}

\def\para#1{\medskip\noindent{\bf#1}}

\begin{document}

\title
{Bayesian analysis of multivariate stochastic volatility with skew distribution}

\author
{{\Large Jouchi Nakajima} \\[1ex]
Department of Statistical Science, Duke University\\
\it jouchi.nakajima@stat.duke.edu
}

\date{December 2012}

\maketitle

\begin{abstract}
Multivariate stochastic volatility models with skew distributions are proposed. Exploiting Cholesky stochastic volatility modeling, univariate stochastic volatility processes with leverage effect and generalized hyperbolic skew t-distributions are embedded to multivariate analysis with time-varying correlations. Bayesian prior works allow this approach to provide parsimonious skew structure and to easily scale up for high-dimensional problem. Analyses of daily stock returns are illustrated. Empirical results show that the time-varying correlations and the sparse skew structure contribute to improved prediction performance and VaR forecasts.\\


\end{abstract}


\section{Introduction}

Multivariate volatility models have attracted attention for their adaptability of variances and correlations to time series dynamics in financial econometrics in particular. A number of works discuss multivariate generalized autoregressive conditional heteroscedasticity (GARCH) models~\citep[see e.g.,][]{BauwensLaurentRombouts06} and multivariate stochastic volatility (MSV) models~\citep[see e.g.,][]{ChibNardariShephard06,AsaiMcAleerYu06,GourierouxJasiakSufana09}. Meanwhile, apart from symmetric distribution, several studies have addressed skew and heavy-tail properties in multivariate financial time series; their modeling strategies for return distributions use skew normal distributions~\citep{AzzaliniDallaValle96,AzzaliniCapitanio99,AzzaliniCapitanio03,Guptaetal04}, a skew-Cauchy distribution~\citep{ArnoldBeaver00}, skew-elliptical distributions~\citep{BrancoDey01,SahuDeyBranco03}, and a finite mixture of skew-normal distributions~\citep{CabralLachosPrates12}. \citep[See][for a survey and discussion of skew distributions for both univariate and multivariate cases]{Azzalini05}. In this context, multivariate GARCH models with skew distributions have been proposed by \cite{BauwensLaurent05} and \cite{AasHaffDimakos06}.

In the literature, little has been discussed about MSV models with skew error distributions. \cite{Zhangetal11} develop a multivariate analysis of the generalized hyperbolic (GH) distribution with time-varying parameters driven by the score of the observation density, based on the generalized autoregressive score (GAS) model~\citep[see also,][]{CrealKoopmanLucas11}. \cite{IshiharaOmoriAsai11} and \cite{IshiharaOmori12} provide MSV models with a leverage effect, a stylized fact of financial returns, which induces skew conditional return distribution. In contrast, the current paper proposes MSV models with leverage effect, where structural errors follow the GH skew $t$-distribution. This is a natural extension of standard univariate stochastic volatility processes with skew distributions~\citep[e.g.,][]{Durham07,SilvaLopesMigon06,NakajimaOmori12} to multivariate analysis; time-varying covariance components are incorporated based on the Cholesky decomposition of volatility matrices, which is increasingly used in time series analysis~\citep[e.g.,][]{PinheiroBates96,SmithKohn02,LopesMcCullochTsay12}. A salient feature is that prior works on a developed Bayesian approach allow for parallel computation of conditional posteriors, which enables the new model to easily scale up to higher dimensions.

Further, the new model includes a structure of skew selection for the multivariate series. Bayesian sparsity modeling has become a popular method to explore parsimonious models in a wide range of statistical analysis~\citep[see e.g.,][]{West2003}. Standard sparsity priors for variable selection in regression models~\citep{GeorgeMcCulloch93,GeorgeMcCulloch97,ClydeGeorge04} are employed for selecting zero or non-zero skewness parameter in the GH skew $t$-distribution for each series. As a related work, \cite{PanagiotelisSmith10} consider the sparsity prior on a coefficient of skew in a multivariate skew $t$-distribution. In the current paper, the sparsity prior is assumed for the skewness parameter in the GH skew $t$-distribution. Empirical studies using time series of stock returns show that the skewness selection, in addition to the dynamic correlated structure, reduces uncertainty of parameters and improves forecasting ability.

Section 2 defines the new MSV models with the GH skew $t$-distribution. Section 3 discusses Bayesian analysis and computation for model fitting. An illustrative example in Section 4 uses a time series of S\&P500 Sector Indices to provide detailed evaluation of the proposed models with comparisons to standard MSV models. Section 5 presents a higher dimensional study of world-wide stock price indices to demonstrate the practical utility of the approach. Section 6 provides some summary comments.


\section{Multivariate stochastic volatility and skew distribution}

This section first introduces the GH skew $t$-distribution in a univariate case in Section \ref{sec:GH}, and then defines the new class of MSV models with the skew distribution in Section \ref{sec:MSV}.


\subsection{GH skew $t$-distribution} \label{sec:GH}

Suppose a univariate time series, $\{w_{it}, t=1,2,\ldots\}$, follows the GH skew $t$-distribution that can be written in the form of \textit{normal variance-mean mixture} as
	\begin{eqnarray}
	w_{it} \,=\, m_i + \beta_i z_{it} + \sqrt{z_{it}}\varepsilon_{it}, \label{eq:w}
	\end{eqnarray}
with $\varepsilon_{it} \sim N(0,1)$, and $z_{it} \sim IG(\nu_i/2,\nu_i/2)$, where $IG$ denotes the inverse gamma distribution. The previous studies often assume that $m_i = -\beta_ic_i$, where $c_i \equiv \mathrm{E}(z_{it}) = \nu_i/(\nu_i-2)$, for $\mathrm{E}(w_{it}) = 0$, and $\nu_i>4$ for the finite variance of $w_{it}$, which is taken here. This is a special case of a more general class of the GH distribution~\citep[see e.g.,][]{AasHaff06}. As discussed by \cite{Prause99} and \cite{AasHaff06}, the parameters of the general GH distribution are typically difficult to jointly estimate. Therefore, the current paper uses the GH skew $t$-distribution in the form of eqn.\,(\ref{eq:w}), which includes necessary parameters enough to describe the skew and heavy-tails of the financial return distributions~\citep{NakajimaOmori12}.

A key structure of the class of GH distributions is that the random variable is represented by the normal variance-mean mixture: a linear combination of two random variables that follow standard normal distribution, and the generalized inverse Gaussian (GIG) distribution, a more general class of the inverse gamma distribution taken for the GH skew $t$-distribution. The combination of a mixing weight $\beta_i$ in eqn.\,(\ref{eq:w}), called an asymmetric parameter, and the scale parameter $\nu_i$ determines the skewness and heavy-tailedness of the resulting distribution. As illustrated by \cite{AasHaff06} and \cite{NakajimaOmori12}, the $\beta_i$ represents the degree of skew with $\nu_i$ fixed, and the $\nu_i$ represents the degree of heavy-tails with $\beta_i$ fixed. There are other definitions for the skew $t$-distributions in the literature~\citep{Hansen94,FernandezSteel98,Prause99,JonesFaddy03,AzzaliniCapitanio03}. However, the GH skew $t$-distribution defined above has a great advantage in consonance with Bayesian modeling of latent variables. The representation of the normal variance-mean mixture leads to an efficient computation with conditional samplers for the latent variables in model fitting using Markov chain Monte Carlo (MCMC) methods, described in Section \ref{sec:Bayes}.


\subsection{Cholesky multivariate stochastic volatility} \label{sec:MSV}

Define a $k\times 1$ vector response time series $\y_t=(y_{1t},\ldots,y_{kt})'$, $(t=1,2,\ldots)$. A standard Cholesky MSV model defines $\y_t \sim N(\zero,\bSigma_t)$ with the triangular reduction $\A_t\bSigma_t\A_t' = \bLambda_t^2$, where $\A_t$ is the lower triangular matrix of covariance components with unit diagonal elements and $\bLambda_t$ is diagonal with positive structural variance elements: viz.
	\begin{eqnarray*}
	\A_t = \left( \begin{array}{cccc}
	1 & 0 & \cdots & 0 \\
	-a_{21,t} & \ddots & \ddots & \vdots \\
	\vdots & \ddots & \ddots & 0 \\
	-a_{k1,t} & \cdots & -a_{k,k-1,t} & 1 \\
	\end{array} \right), \quad
	\bLambda_t = \left( \begin{array}{cccc}
	\lambda_{1t} & 0 & \cdots & 0 \\
	0 & \ddots & \ddots & \vdots \\
	\vdots & \ddots & \ddots & 0 \\
	0 & \cdots & 0 & \lambda_{kt} \\
	\end{array} \right).
	\end{eqnarray*}
This implies $\bSigma_t = \A_t^{-1}\bLambda_t^2(\A_t')^{-1}$, and $\y_t=\A_t^{-1}\bLambda_t\e_t$, where $\e_t\sim N(\zero,\I)$. The construction of this Cholesky decomposition has appeared in previous works for constant covariance matrices~\citep{PinheiroBates96,Pourahmadi99,SmithKohn02,GeorgeSunNi08} and dynamic covariance modeling with stochastic volatility models~\citep{CogleySargent05,Primiceri05,LopesMcCullochTsay12}. A salient feature in Bayesian modeling of Cholesky MSV models for time-varying parameters of the covariance/variance elements is that the approach reduces the multivariate dynamics to univariate volatility processes that form a state space representation, as discussed by \cite{LopesMcCullochTsay12}. The new idea exploits the Cholesky structure for modeling MSV and embeds the GH skew $t$-distribution as follows.

The new class of models is defined by
	\begin{eqnarray*}
	\y_t = \A_t^{-1}\bLambda_t\w_t,
	\end{eqnarray*}
where $\w_t = (w_{1t},\ldots,w_{kt})'$ is the $k\times 1$ vector whose element $w_{it}$ independently follows the GH skew $t$-distribution defined by eqn.\,(\ref{eq:w}). Define $\h_t = (h_{1t},\ldots,h_{kt})'$ as the $k\times 1$ vector of stochastic volatility in $\bLambda_t$ with $h_{it} = \log(\lambda_{it}^2)$, for $i=1,\ldots,k$, and $\a_t = (a_{1t},\ldots,a_{pt})'$ as the $p\times 1$ ($p=k(k-1)/2$) vector of the strictly lower-triangular elements of $\A_t$ (stacked by rows). The time-varying processes for these Cholesky parameters are specified as
	\begin{eqnarray}
	\h_{t+1} &=& \bmu + \bPhi(\h_t - \bmu) + \bfeta_t, \nonumber \\
	\a_{t+1} &=& \bmu_a + \bPhi_a(\a_t - \bmu_a) + \bxi_t, \label{eq:at}
	\end{eqnarray}
and
	\begin{eqnarray*}
	\left( \begin{array}{c} \beps_t \\ \bfeta_t \\ \bxi_t \end{array} \right) &\sim& N \left( \zero, \left[
	\begin{array}{ccc} \I & \Q & \O \\ \Q & \S & \O \\ \O & \O & \V \end{array} \right] \right),
	\end{eqnarray*}
where $\beps_t = (\varepsilon_{1t},\ldots,\varepsilon_{kt})'$, and each of $(\bPhi,\bPhi_a,\S,\Q,\V)$ is assumed diagonal: $\bPhi=\diag(\phi_i)$, $\bPhi_a=\diag(\phi_{ai})$, $\S=\diag(\sigma_i^2)$, $\Q=\diag(\rho_i\sigma_i)$, and $\V=\diag(v_{ai}^2)$, with $|\phi_i|<1$ and $|\phi_{aj}|<1$, for each $i=1,\ldots,k$, and $j=1,\ldots,p$. Thus all univariate time-varying parameters follow stationary AR(1) processes. The identity $\tilde{\y}_t \equiv \A_t\y_t = \bLambda_t \w_t$ leads to a set of univariate stochastic volatilities with the GH skew $t$-distribution~\citep{NakajimaOmori12}:
	\begin{eqnarray}
	\tilde{y}_{it} &=& \{ \beta_i(z_{it}-c_i) + \sqrt{z_{it}}\varepsilon_{it} \} \exp(h_{it}/2), \label{eq:sv_yt} \\ 
	h_{i,t+1} &=& \mu_i + \phi_i(h_{it} - \mu_i) + \eta_{it}, \\
	\left( \begin{array}{c} \varepsilon_{it} \\ \eta_{it} \\ \end{array} \right)
	&\sim& N(\zero,\bOmega_i),
	\quad\mathrm{and}\quad
	\bOmega_i=
	\left( \begin{array}{cc} 1 & \rho_i\sigma_i \\ \rho_i\sigma_i & \sigma_i^2 \\ \end{array} \right), \label{eq:sv_S}
	\end{eqnarray}
where $(\tilde{y}_{it},\mu_i,\phi_i,\eta_{it})$ are the $i$-th (diagonal) elements of $(\tilde{\y}_t, \bmu, \bPhi, \bfeta_t)$, respectively, for $i=1,\ldots,k$. The $\rho_i$ measures the correlation between $\varepsilon_{it}$ and $\eta_{it}$, which is typically negative for stock returns as the so-called leverage effect~\citep{Yu05,OmoriChibShephardNakajima07}. The class of univariate stochastic volatility models has been well studied in the literature~\citep[e.g.,][]{JacquierPolsonRossi04,KimShephardChib98,GhyselsHarveyRenault02,Eraker04,Shephard05,NakajimaOmori09}.

In the context of MSV modeling~\citep{ChibNardariShephard06,AsaiMcAleerYu06,GourierouxJasiakSufana09}, the proposed model here is a natural extension of the univariate stochastic volatility model with the GH skew $t$-distributions embedded in the Cholesky-type multivariate structure. Most of the multivariate skew distributions and their extension to volatility models in the previous literature often have the difficulty of scaling up in dimension of responses in terms of computation of model likelihoods and parameter estimates. In contrast, an inference of the new model reduces to that of simply $k$ univariate stochastic volatility models; this leads an efficient and fast parallel computation under conditionally independent priors as specified below.


\subsection{Skew selection} \label{sec:sparse}

As mentioned by \cite{Primiceri05}, it is not straightforward to theoretically explore compounded processes of covariance/variance elements in the Cholesky-type covariance matrix. (See Appendix B of \cite{Nakajimaphd12} for characteristics of the resulting covariance matrix process.) To understand the skew in the Cholesky MSV models, a simulation study follows.

\begin{figure}[t]

\centering
\includegraphics[width=4.2in]{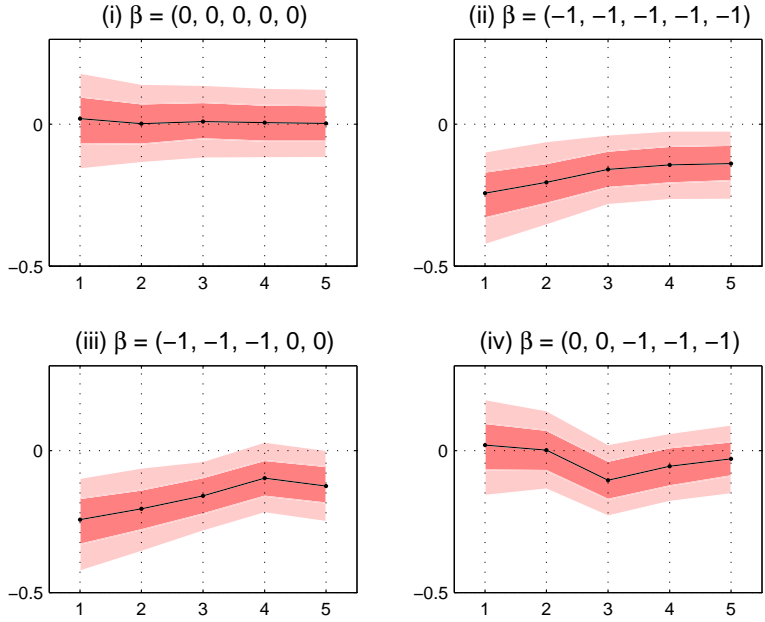} 

\caption{Skewness of simulated data: means (solid line), 50\% (filled area, dark) and 80\% (light) intervals for 1{,}000 sets of simulated series. The horizontal axis refers to the series index $i$.}

\label{fig-sim}
\end{figure}

A sample of size $T=1{,}000$ and $k=5$ is simulated according to the proposed MSV model with fixed parameter values: $\phi_i=0.995$, $\sigma_i=0.05$, $\rho_i=-0.5$, $\mu_i=-9$, $\nu_i=20$, and $a_j = 0.5$, for all $i,j$. These values are selected following empirical studies in previous works. The value of $a_j$'s set here implies correlations between responses around 0.3--0.8. For the skewness parameter, four sets of values are considered: $\bbeta \equiv (\beta_1,\ldots,\beta_k)=$ (i) $\zero_{1 \times 5}$, (ii) $-\one_{1\times 5}$, (iii) $(-1,-1,-1,0,0)$, and (iv) $(0,0,-1,-1,-1)$. Figure \ref{fig-sim} shows summaries of skewness of simulated data from $1{,}000$ sets of simulation. The cases (i) and (ii) clearly exhibit no skewness and significant skewness, respectively. Interestingly, the case (iii) still yields skew observations including in the last two series ($i=4,5$) despite the zero skewness parameters. This is because the latter series inherit the former structural processes due to the lower triangular structure of Cholesky components (see Appendix B of \cite{Nakajimaphd12}). The case (iv) confirms this mechanism; the first two series do not exhibit skewness because the corresponding skewness parameters are zero, and no inherited structure arises.

From these findings, the skewness parameter $\beta_i$'s can be redundant for the latter series in the response vector $\y_t$. Shrinkage to zero of subsets of the skewness parameters addresses skew selection in the Cholesky MSV model, exploring more parsimonious structure to reduce estimation uncertainty and improve predictions. A traditional sparsity prior for variable selection in regression models~\citep{GeorgeMcCulloch93,GeorgeMcCulloch97,ClydeGeorge04} is employed for the skew selection. Specifically, the sparsity prior for $\beta_i$ has the form
	\begin{eqnarray}
	\beta_i \,\sim\, \kappa N(\beta_i|0, \tau_0^2) + (1-\kappa)\delta_0(\beta_i), \label{eq:beta}
	\end{eqnarray}
for $i=1,\ldots,k$, where $\delta_0$ denotes the Dirac delta function at zero. This prior assigns the probability $\kappa$ of taking a non-zero value and the shrinkage probability $1-\kappa$ with a point mass at zero. Due to the structure of the normal-mean variance mixture and the conditional independence of univariate stochastic volatility processes, a conditional sampler for $\beta_i$ under the sparsity prior is quite easy and simple as described in the next section.


\section{Bayesian analysis and computation} \label{sec:Bayes}

Model fitting using the MCMC methods includes conditional samplers for univariate stochastic volatility models with leverage effect (\cite{OmoriChibShephardNakajima07}; \cite{OmoriWatanabe08}; \cite{NakajimaOmori12}) and for the state space dynamic models~\citep[e.g.,][]{PradoWest10}. Based on observations $\y_{1:T} = \{\y_1,\ldots,\y_T\}$ over a given time period of $T$ intervals, the full set of latent process state parameters and model parameters in the posterior analysis are listed as follows:
	\begin{itemize}
	\item The stochastic volatility processes $h_{i,1:T}$ and mixing latent processes $z_{i, 1:T}$, $(i=1,\ldots,k)$;
	\item The covariance component process states $\a_{1:T}$;
	\item The skewness parameters $\bbeta$ and the sparsity hyper-parameter $\kappa$;
	\item Hyper-parameters defining each of the univariate stochastic volatility processes, $\btheta_i \equiv \{\mu_i,\phi_i,\sigma_i,\rho_i,\nu_i\}$, $(i=1,\ldots,k)$;
	\item Hyper-parameters defining each of the covariance component processes, $\btheta_{aj} \equiv \{\mu_{aj},\phi_{aj},$ $v_{aj}\}$, $(j=1,\ldots,p)$.
	\end{itemize}
Components of the MCMC computations are outlined as follows.

\para{Stochastic volatility processes and mixing latent processes:}
The conditional posteriors for each of the latent volatility processes $h_{i,1:T}$, $(i=1,\ldots,k)$ are sampled using the MCMC technique for the stochastic volatility models with leverage~\citep{OmoriChibShephardNakajima07, OmoriWatanabe08}. \cite{NakajimaOmori12} implement the algorithm for the stochastic volatility  with the GH skew $t$-distribution. Including the mixing latent process, $z_{i, 1:T}$, $(i=1,\ldots,k)$, these state processes are conditionally independent across $i$ in the posteriors given all other latent variables and hyper-parameters, which allows parallel generation of the volatility processes based on eqns.\,(\ref{eq:sv_yt})-(\ref{eq:sv_S}).

\para{Covariance component process states:}
Conditional on other latent process states and hyper-parameters, the MSV model reduces to a conditionally linear, Gaussian dynamic model for the states $\a_{1:T}$. Specifically, 
	\begin{eqnarray*}
	\hat{y}_{it} &=& \a_t^{(i)}\x_{it} + \hat{\varepsilon}_{it},\quad
	\hat{\varepsilon}_{it} \sim N\left(0, z_{it}(1-\rho_i^2)\right),
	\end{eqnarray*}
where $\hat{y}_{it}= y_{it}e^{-h_t/2}-\beta_i(z_{it}-c_i)-\sqrt{z_{it}}\rho_i\hat{\eta}_{it}$, $\hat{\eta}_{it}=(h_{i,t+1}-\mu_i) - \phi_i(h_{it}-\mu_i)$, $\x_{it} = (y_{1t}e^{-h_{1t}/2}, \ldots,$ $y_{i-1,t}e^{-h_{i-1,t}/2})'$, and  $\a_t^{(i)}$ denotes the $1\times (i-1)$ vector of the free parameters in the $i$-th row of $\A_t$, for $i=2,\ldots,k$. This observation equation is coupled with the state evolution of eqn.\,(\ref{eq:at}); sampling full sets of the states is implemented using the standard forward filtering, backward sampling (FFBS) algorithm~\citep[e.g.,][]{deJongShephard95}.

\para{Skewness parameters:}
Conditional on all the latent states and hyper-parameters, under the prior defined by eqn.\,(\ref{eq:beta}), the posterior for the skewness parameter $\beta_i$ is given by
	\begin{eqnarray*}
	\beta_i \,|\, \cdot \,\sim\, \hat{\kappa}_i N(\beta_i|\hat{\beta}_i, \hat{\tau}_i^2) + (1-\hat{\kappa}_i)\delta_0(\beta_i),
	\end{eqnarray*}
where $\hat{\beta}_i$ and $\hat{\tau}_i^2$ are the posterior mean and variance of the posterior distribution for $\beta_i$ under the normal prior $N(0,\tau_0^2)$; and $\hat{\kappa}_i = \kappa b_i/(\kappa b_i+1-\kappa)$ with $b_i = \exp(\hat{\beta}_i^2/2\hat{\tau}_i^2)\hat{\tau}_i/\tau_0$. For the parameter $\kappa$, a beta prior is assumed; then we directly sample the conditional posterior given the number of $\beta_i$'s such that $\beta_i \neq 0$.

\para{Stochastic volatility hyper-parameters:}
For each $i=1,\ldots,k$, traditional forms of priors for AR model parameters are assumed: normal priors for $\mu_i$, shifted beta priors for each of ($\phi_i,\rho_i$), inverse gamma priors for $\sigma_i^2$, and truncated gamma priors for $\nu_i$ ($\nu_i>4$), with prior independence across $i$. Conditional posteriors, given the other state variables and hyper-parameters, can be sampled directly or via Metropolis-Hastings algorithms. \citep[See][Section 2.2]{NakajimaOmori12}.

\para{AR hyper-parameters $\btheta_{aj}$:}
For each $j=1,\ldots,p$, the same forms of priors are assumed for $(\mu_{aj},\phi_{aj}, v_{aj}^2)$ as $(\mu_i,\phi_i,\sigma_i^2)$. Conditional posteriors given the states $\a_{1:T}$ can be sampled directly or via Metropolis-Hastings algorithms.

\medskip
Note that sampling each of $(h_{i,1:T}, z_{i,1:T}, \a_{1:T}^{(i)}, \beta_i, \btheta_i)$ can be parallelized across $i$. In preliminary simulation studies and the following empirical examples, MCMC streams were fairly clean and stable with quickly decaying sample autocorrelations in the same manner as the univariate stochastic volatility models.


\section{A study of stock price index} \label{sec:sp}

The first study applies the proposed model to a series of $k=5$ daily stock returns. An analysis particularly focuses on how the multivariate correlation mechanism and skew components reveal dynamic relationships underlying the stock return volatilities and improve forecasting ability. Note some connections with previous work on multivariate stock return time series using dynamic volatility models~\citep[e.g.][]{AasHaffDimakos06,ChibNardariShephard06,ConradKaranasosZeng11,Zhangetal11,IshiharaOmori12}.


\subsection{Data and model setup} \label{sec:data}

\begin{table}[t]
\centering
\begin{tabular}{cll}

\hline

1	&	INDU	&	Industrials	\\
2	&	CONS	&	Consumer Staples \\
3	&	FINL	&	Financials	\\
4	&	ENRS	&	Energy	\\
5	&	INFT	&	Information Technology \\

\hline

\end{tabular}

\caption{S\&P500 Sector Index. Sectors are ordered by smaller posterior means of the skewness parameter $\beta_i$ obtained from univariate stochastic volatility models with the skew $t$-distribution.}
\label{tab-list}
\end{table}

The data are S\&P500 Sector Indices over a time period of 1{,}510 business days beginning in January 2006 and ending in December 2011. The returns are computed as the log difference of the daily closing price. The series are listed in Table \ref{tab-list}. The ordering of the series in the vector of response $\y_t$ matters due to the structure based on the Cholesky decomposition. From simulation results in Section \ref{sec:sparse}, the series are ordered by smaller posterior means of the skewness parameter $\beta_i$ obtained from the univariate stochastic volatility models with the skew $t$-distribution. This strategy induces more parsimonious skew structure, which improves forecasting performance as discussed below.

The following priors are used: $(\phi_*+1)/2 \sim B(20,1.5)$ for each $*\in \{i, ai\}$, $\mu_i \sim N(-10,1)$, $\mu_{ai} \sim N(0,1)$, $\sigma_i^{-2} \sim G(20, 0.01)$, $v_{ai}^{-2} \sim G(20, 0.01)$, $(\rho_i+1)/2 \sim B(1, 1)$, $\nu_i \sim G(16, 0.8)I[\nu_i>4]$, $\beta_i \sim \kappa N(\beta_i|0, 10) + (1-\kappa)\delta_0(\beta_i)$, and $\kappa \sim B(2, 2)$, where $B$ and $G$ denotes the beta and gamma distributions, respectively. The MCMC analysis was run for a burn-in period of 5{,}000 samples prior to saving the following 50{,}000 samples for posterior inferences.

The study provides forecasting performance in comparison among different specifications in the proposed class of models. The following five models are considered:
	\begin{itemize}
	\item {\bf S:} Skew $t$-distribution, no sparsity on $\beta_i$ ($\kappa \equiv 1$), no correlation ($\A_t \equiv \I$);
	\item {\bf SS:} Skew $t$-distribution with sparsity on $\beta_i$, no correlation;
	\item {\bf C:} Symmetric $t$-distribution ($\beta_i \equiv 0$), with correlation;
	\item {\bf CS:} Skew $t$-distribution, no sparsity on $\beta_i$, with correlation;
	\item {\bf CSS:} Skew $t$-distribution with sparsity on $\beta_i$, and correlation.
	\end{itemize}
The key focus here is on the skew in return distribution, sparsity structure on the skewness parameter, and the Cholesky-type correlation mechanism in the MSV.


\subsection{Forecasting performance and comparisons} \label{sec:forecast}

Out-of-sample forecast performance is examined to compare the competing models in predicting 1 to 5 business days ahead. Forecasts are based on a posterior predictive density sampled every MCMC iteration. An experiment is implemented in a traditional recursive forecasting format; the full MCMC analysis is fit to each model to obtain the 5-horizon forecasts given data from the start of January 2006 up to business day $T_i$ with $T_i = T_{i-1} + 5$. Specifically, each model is first estimated based on data $\y_{1:T_1}$ where $T_1=1{,}010$. The resulting out-of-sample predictive distributions are simulated over the following 5 business days, $t=T_1+1,\ldots,T_1+5$. Next, the analysis moves ahead 5 business days to observe the next 5 observations $\y_{T_1+1:T_1+5}$ and reruns the MCMC based on the updated data $\y_{1:T_2}$, where $T_2 = T_1+5$, forecasting the following 5 business days $t=T_2+1,\ldots,T_2+5$. This is repeated with $T_i=T_{i-1}+5$ for $i=2,\ldots,100$, generating a series of out-of-sample forecasts over 500 business days. This experiment allows us to explore forecasting performance over nearly 2-year periods of different financial market circumstances and so examine robustness to time periods of the prediction ability.

\begin{table}[b]
\centering
\begin{tabular}{lrrrrrr}

\hline
		&	\multicolumn{5}{c}{Horizon ($d$ days)}	&	\\
Model &	 1     &  2     &  3     &  4	 &  5      & Total \\
\hline
SS    &   7.0  &  13.2  &  11.2  &  16.7 &   15.4  &  63.5 \\
C     &  34.0  &  54.7  &  45.5  &  52.8 &   50.9  & 237.9 \\
CS    &  88.5  & 216.8  &  85.2  & 240.2 &  144.5  & 775.1 \\
CSS   &  92.8  & 230.6  &  91.5  & 259.7 &  156.9  & 831.4 \\

\hline

\end{tabular}

\caption{Cumulative log predictive density ratios $\mathrm{LPDR}_t(d)$ relative to Model S.}
\label{tab-dens}
\end{table}

The first measure of formal model assessments is out-of-sample predictive densities. The log predictive density ratio (LPDR) for forecasting $d$ business days ahead from the day $t$ is $\textrm{LPDR}_t(d) = \log \{ p_{M_1}(\y_{t+d}|\y_{1:t}) / p_{M_0}(\y_{t+d}|\y_{1:t})\}$, where $p_M(\y_{t+d}|\y_{1:t})$ is the predictive density under model $M$. This quantity represents relative forecasting accuracy in the prediction exercise. Table \ref{tab-dens} reports the LPDRs of four competing models relative to Model S at each horizon. Improvements in out-of-sample predictions are practically evident for the proposed multivariate skew models. The LPDRs for Models C and CS show relevance of correlated structure, and differences in those for Models C and CS indicate dominance of the skew component in the multivariate stock returns. The LPDRs for Model SS and comparisons in those for Models CS and CSS show that the sparsity on the skew parameters contributes to improved predictions, robustly across horizons. The LPDRs for Models CS and CSS at the 2nd and 4th horizons are relatively inflated, which is due to two time points under market shocks related to the European sovereign-debt crisis. In turbulent situations, the skew and correlated structures yield substantially increased improvements. Even if these two times are removed from the full period of comparison, the models still show relevant dominance over the standard MSV models.

The second measure of the formal model comparisons is based on Value-at-Risk (VaR) forecasts of portfolio returns. Using samples from the posterior predictive distribution, optimal portfolios are implemented under several allocation rules, and the VaR forecast of the resulting portfolio is obtained at each time $T_i+1,\ldots,T_i+5$, for $i=1\ldots,100$. Note that \cite{BauwensLaurent05} illustrate a similar procedure in evaluation of the VaR forecasts. A main focus here is on an impact of the proposed multivariate skew model on forecasting accuracy, in particular for a tail risk of multivariate responses.

The analysis uses standard Bayesian mean-variance optimization~\citep{Markowitz59}. Based on the samples from the posterior predictive distribution, the forecast mean vector and variance matrix of $\y_t$, denoted by $\g_t$ and $\D_t$ respectively, are computed. Investments are allocated according to a vector of portfolio weights, denoted by $\bomega_t$, optimized by the following allocation rule. The realized portfolio return at time $t$ is $\bomega_t'\y_t$. Given a (scalar) return target $m$, we optimize the portfolio weights $\bomega_t$ by minimizing the forecast variance of the portfolio return among the restricted portfolios whose expectation is equal to $m$. Specifically, we minimize an ex-ante portfolio variance $\bomega_t'\D_t\bomega_t$, subject to $\bomega_t'\g_t = m$, and $\bomega_t'\one = 1$, i.e., the total sum invested on each business day is fixed. The solution is $\bomega_t^{(m)} = \K_t (\one' \K_t \q_t \g_t - \g_t' \K_t \q_t \one)$, where $\q_t = (\one m - \g_t) / d_t$, and $d_t = (\one' \K_t \one)(\g_t' \K_t \g_t) - (\one' \K_t \g_t)^2$, with $\K_t = \D_t^{-1}$. The study also considers the target-free minimum-variance portfolio given by $\bomega_t^* = \K_t \one / (\one'\K_t\one)$.  The portfolio is reallocated on each business day based on 1- to 5- business day ahead forecasts. This experiment assumes a practical situation that investors allocate their resource every business day based on weekly-updated forecasts. Note that the resources are assumed freely reallocated to arbitrary long or short positions without any transaction cost.

\begin{table}[p]
\centering
\begin{tabular}{lrrrrr}

\multicolumn{6}{l}{(1) Violations} \\
\hline
	  &  &	\multicolumn{3}{l}{Target return ($m$)}	&	Target \\
Model &	\multicolumn{1}{c}{$\alpha$} &	0.005\%	&	0.01\%	&	0.02\%	& -free \\
\hline

S     & 0.5\% &  27   &  18  &  12  &  37 \\
      & 1\%   &  37   &  26  &  12  &  50 \\
      & 5\%   &  67   &  53  &  30  &  79 \\
SS    & 0.5\% &  31   &  20  &  7   &  41 \\
      & 1\%   &  43   &  32  &  11  &  48 \\
      & 5\%   &  69   &  60  &  41  &  79 \\
C     & 0.5\% &  30   &  22  &  12  &  45 \\
      & 1\%   &  39   &  29  &  18  &  52 \\
      & 5\%   &  64   &  48  &  33  &  82 \\
CS    & 0.5\% &  2    &  2   &  2   &  5  \\
      & 1\%   &  5    &  7   &  3   &  8  \\
      & 5\%   &  18   &  19  &  17  &  25 \\
CSS   & 0.5\% &  3    &  2   &  1   &  4  \\
      & 1\%   &  6    &  5   &  2   &  6  \\
      & 5\%   &  24   &  26  &  22  &  24 \\

\hline

\\

\multicolumn{6}{l}{(2) $p$-values} \\
\hline
	  &  &	\multicolumn{3}{l}{Target return ($m$)}	&	Target \\
Model &	\multicolumn{1}{c}{$\alpha$} &	0.005\%	&	0.01\%	&	0.02\%	& -free \\
\hline

S     & 0.5\% &  0.00  &  0.00  &  0.00  &  0.00 \\
      & 1\%   &  0.00  &  0.00  &  0.01  &  0.00 \\
      & 5\%   &  0.00  &  0.00  &  0.32  &  0.00 \\
SS    & 0.5\% &  0.00  &  0.00  &  0.02  &  0.00 \\
      & 1\%   &  0.00  &  0.00  &  0.02  &  0.00 \\
      & 5\%   &  0.00  &  0.00  &  0.00  &  0.00 \\
C     & 0.5\% &  0.00  &  0.00  &  0.00  &  0.00 \\
      & 1\%   &  0.00  &  0.00  &  0.00  &  0.00 \\
      & 5\%   &  0.00  &  0.00  &  0.12  &  0.00 \\
CS    & 0.5\% &  0.74  &  0.74  &  0.74  &  0.16 \\
      & 1\%   &  1.00  &  0.40  &  0.33  &  0.22 \\
      & 5\%   &  0.13  &  0.20  &  0.08  &  1.00 \\
CSS   & 0.5\% &  0.76  &  0.74  &  0.28  &  0.38 \\
      & 1\%   &  0.66  &  1.00  &  0.13  &  0.66 \\
      & 5\%   &  0.84  &  0.84  &  0.53  &  0.84 \\
\hline

\end{tabular}

\caption{VaR results: the number of violations and $p$-values for the likelihood ratio test. The null hypothesis is that the expected ratio of violations is equal to $\alpha$.}
\label{tab-test}
\end{table}

In summary of the VAR forecasts, the number of VaR violations, denoted by $n$, is counted over $N=500$ experiment days. The expected number of violations for $\alpha$ quantile is $\alpha N$; under the null hypothesis that the expected ratio of violations is equal to $\alpha$, the likelihood ratio statistic,
	\begin{eqnarray*}
	2\log \left\{ \left(\frac{n}{N}\right)^n \left(1-\frac{n}{N}\right)^{N-n} \right\} - 2\log\left\{ \alpha^n (1 - \alpha)^{N-n} \right\},
	\end{eqnarray*}
is asymptotically distributed as $\chi^2(1)$ \citep[see][]{Kupiec95}. Table \ref{tab-test} reports the number of VaR violations and results of the likelihood ratio test for $\alpha=0.5\%$, $1\%$, and $5\%$ levels, based on a range of daily target returns of $m=0.005\%$, $0.01\%$, and $0.02\%$, implying a yearly return of approximately $1.25\%$, $2.5\%$, and $5.0\%$ respectively, as well as the target-free portfolio. For a 5\% significance level, the null hypothesis is rejected in almost all cases of VaR quantiles and portfolio rules for Models S, SS, and C. The large number of their VaR violations indicate that these models forecast smaller values of VaR (in their absolute value) than necessary. This optimistic risk forecast is due to lack of structure including both skewness and correlations. In contrast, the null hypothesis is not rejected in all cases for Models CS and CSS except in only one case, $\alpha=5\%$ with $m=0.02\%$ for Model CS.

The results from the out-of-sample forecasting experiments reveal that the skew and correlated multivariate structure contributes to the forecasting performance in terms of the predictive density and the VaR risk analysis. In particular, the sparse skew model with correlation, Model CSS, achieved the best posterior predictive densities and passed the VaR likelihood ratio test for all the realistic situations. These findings are similar to those obtained from different prior densities; a prior sensitivity analysis is provided in the Appendix.

Regarding the ordering of the response in $\y_t$, Bayesian prior works on the orderings can be considered~\citep[see][for reversible-jump MCMC methods in the Cholesky MSV models]{NakajimaWatanabe11}, although this is beyond the scope of this paper. Instead, the reverse ordering of the responses is examined here to compare with the current baseline ordering. Forecasting performances are computed for Model CSS; results show weaker forecasting ability of the reverse ordering than the baseline ordering in terms of the predictive density and VaR forecasts. This confirms that the baseline ordering based on the posterior means of $\beta_i$'s has an advantage over the reverse ordering.


\subsection{Summaries of posterior inferences}

\begin{figure}[p]

\centering
\includegraphics[width=4.9in]{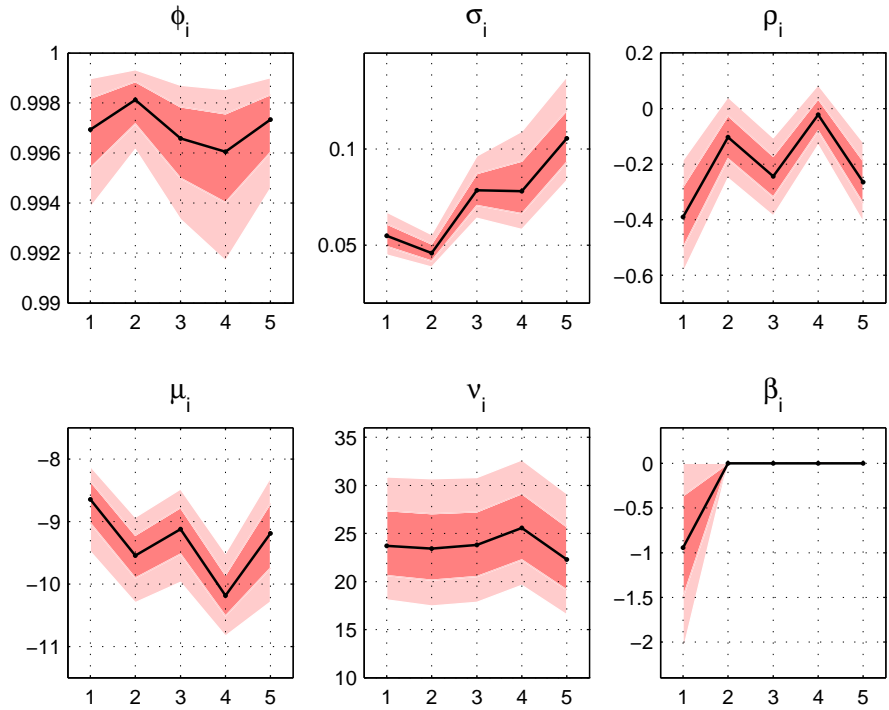} 

\caption{Posterior estimates for parameters from Model CSS: Posterior medians (solid line) and 50\% (filled area, dark) and 80\% (light) credible intervals. The horizontal axis refers to the series index $i$.}

\label{fig-par}
\end{figure}

\begin{figure}[p]

\centering
\includegraphics[width=4.1in]{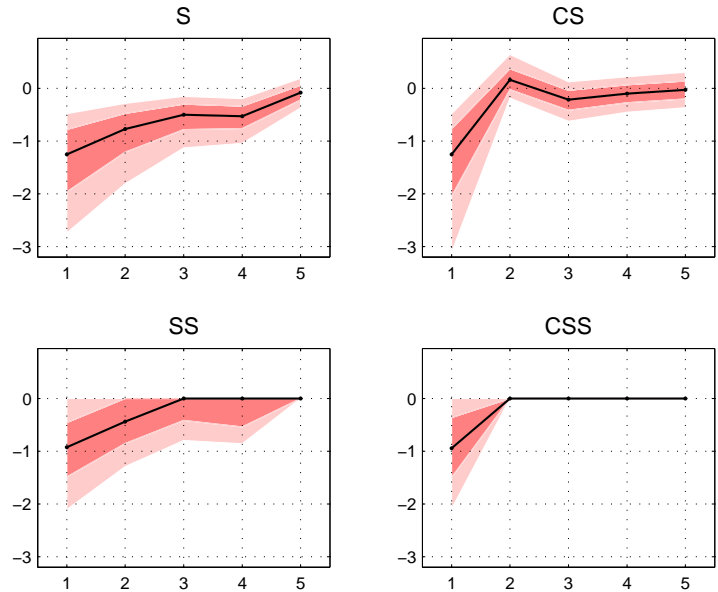}

\caption{Posterior estimates for $\beta_i$: Posterior medians (solid line) and 50\% (filled area, dark) and 80\% (light) credible intervals. The horizontal axis refers to the series index $i$.}

\label{fig-beta}
\end{figure}

Posterior estimates are summarized for results of the MSV models fit to data $\y_{1:T}$ with $T=T_1=1{,}010$. Figure \ref{fig-par} displays posteriors of model parameters for Model CSS. One remarkable finding is that the posterior for $\beta_1$ is estimated negative considerably apart from zero, although the posteriors for other $\beta_i$'s ($i=2,\ldots,5$) exhibit shrinkage at zero; their posterior probabilities of shrinkage are about 91--94\%. This parsimonious skew structure evidently improves forecasting ability compared to the non-sparsity model as reported in the previous subsection. Figure \ref{fig-beta} plots the posterior estimates of $\beta_i$ for four competing models. For Models S the $\beta_i$'s are estimated negative with reported credible intervals that are mostly apart from zero. For Model SS, moderate shrinkages are found for ($\beta_3, \beta_4$), and considerable shrinkage is observed for $\beta_{5t}$. In contrast, Model CS exhibits credible intervals including zero except for $\beta_1$, and the evident shrinkages in Model CSS yield the parsimonious skew structure.

The posteriors for the other parameters in Figure \ref{fig-par} are consistent with previous studies. The posterior medians of $\rho_i$'s are estimated negative, indicating the leverage effect for stock return dynamics. One possible extension from the current model is sparsity for $\rho_i$'s, although a result from Model CSS with the same sparsity prior embedded to $\rho_i$'s showed only little evidence of sparsity for all $i$.

\begin{figure}[t]

\centering
\includegraphics[width=4.8in]{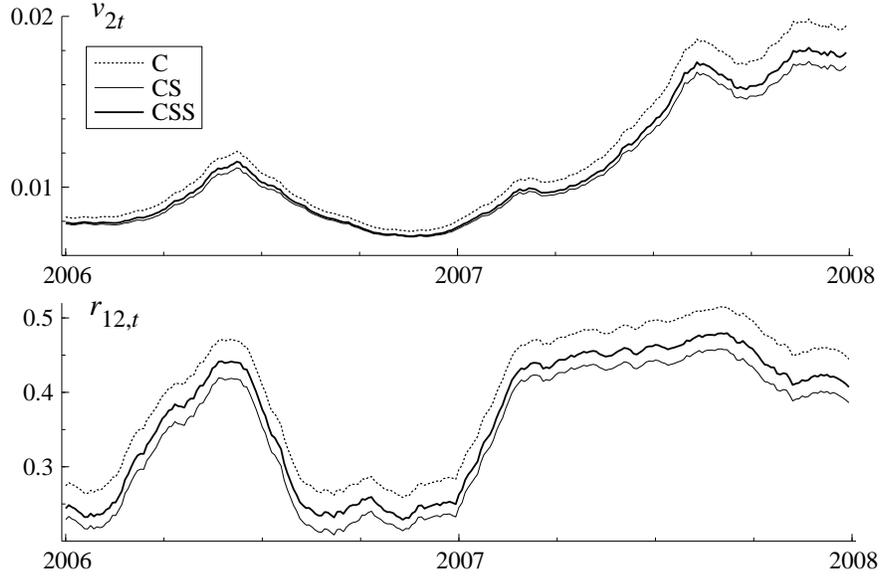} 

\caption{Posterior means of the selected standard deviation (top, CONS) and correlation (bottom, INDU-CONS) in $\bSigma_t$.}

\label{fig-sig}
\end{figure}

Figure \ref{fig-sig} graphs trajectories of posterior means of a selected standard deviation and correlation, denoted by $v_i$ and $r_{ij}$ respectively, in the resulting covariance matrix $\bSigma_t  = \A_t^{-1}\bLambda_t^2(\A_t')^{-1}$. Note that the figure shows only a part of sample periods for visual clarity. The top panel shows the standard deviation of CONS ($i=2$) from three MSV models. Model C yields higher standard deviations than the other skewed models due to the symmetric $t$-distribution that estimates the left tail lighter than the skew models. Model CSS yields higher standard deviations than Model CS because of the shrinkage toward zero. These differences tend to be larger in high-volatility periods. The same feature is found in the correlations; the bottle panel shows the correlation between INDU ($i=1$) and CONS. Model CS yields less correlated structure due to its skew error distribution. Meanwhile, across the series and sample periods, the correlation is evidently time-varying for the stock return data, which results in the contribution of the Cholesky-based time-varying correlation structure to the improved prediction ability.

\begin{figure}[t]

\centering
\includegraphics[width=5.5in]{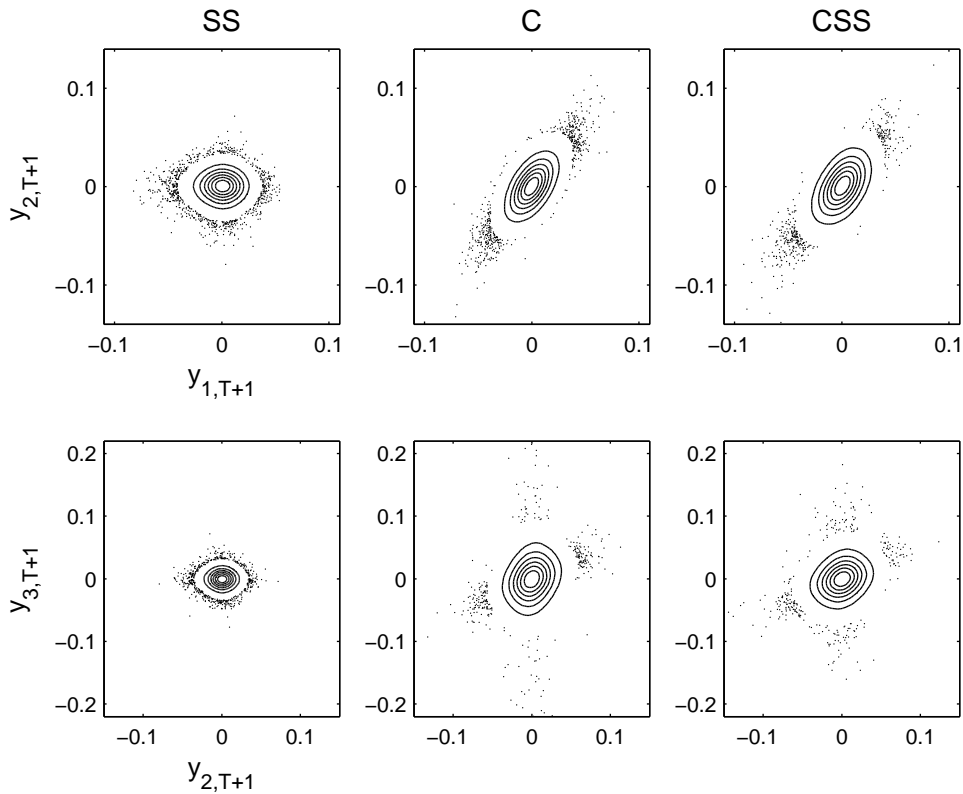} 

\caption{Surface plots for smoothed joint predictive density for ($y_{1,T+1},y_{2,T+1}$) (top) and ($y_{2,T+1},y_{3,T+1}$) (bottom) based on MCMC samples. Scatter plots are displayed  for tail samples, defined as samples in regions where the smoothed density is less than 1\% of the maximum density.}

\label{fig-scat}
\end{figure}

Further, Figure \ref{fig-scat} shows approximated posterior joint predictive densities of ($y_{1,T+1},y_{2,T+1}$) and ($y_{2,T+1},y_{3,T+1}$) in surface plots with tail behaviors displayed in scatter plots. Compared to Model SS, the correlated MSV models (C and CSS) exhibit a clear image of correlated predictive densities. Model CSS yields more tail samples in the left tails due to the negative skewness. These differences result in the large improvement of VaR forecasts illustrated in the previous subsection.


\section{A higher-dimensional study: World-wide stock price indices}

\begin{table}[t]
\centering
\begin{tabular}{llll}

\hline

1	&	Euro		 &	11 &	Brazil		\\
2	&	US			 &	12 &	Spain		\\
3	&	India		 &	13 &	Russia		\\
4	&	Taiwan		 &	14 &	Swiss		\\
5	&	Netherlands	 &	15 &	Hong Kong	\\
6	&	Japan		 &	16 &	UK			\\
7	&	Mexico		 &	17 &	Australia	\\
8	&	Sweden		 &	18 &	Germany		\\
9	&	France		 &	19 &	Canada		\\
10	&	Italy		 &	20 &	Korea		\\

\hline

\end{tabular}

\caption{World-wide stock price indices. Countries are ordered by smaller posterior means of the skewness parameter $\beta_i$ obtained from univariate stochastic volatility models with the skew $t$-distribution.}
\label{tab-list-w}
\end{table}

This section provides a higher-dimensional example for the skew and correlated MSV models using $k=20$ world-wide stock price indices (see the list of countries and regions in Table \ref{tab-list-w}). These are selected as major indices traded in the global financial market; note that both the Euro and several European countries are included, although their time series do not exhibit severely high correlations. The time period is $T=1{,}258$ business days beginning in January 2006 and ending in December 2010. The returns are computed as the log difference of prices at the closing time of the US market. The variables in $\y_t$ are ordered by posterior means of the skewness parameter $\beta_i$ obtained from the same pre-analysis, and the study uses the same prior specifications as in the previous section.

\begin{figure}[p]

\centering
\includegraphics[width=5.8in]{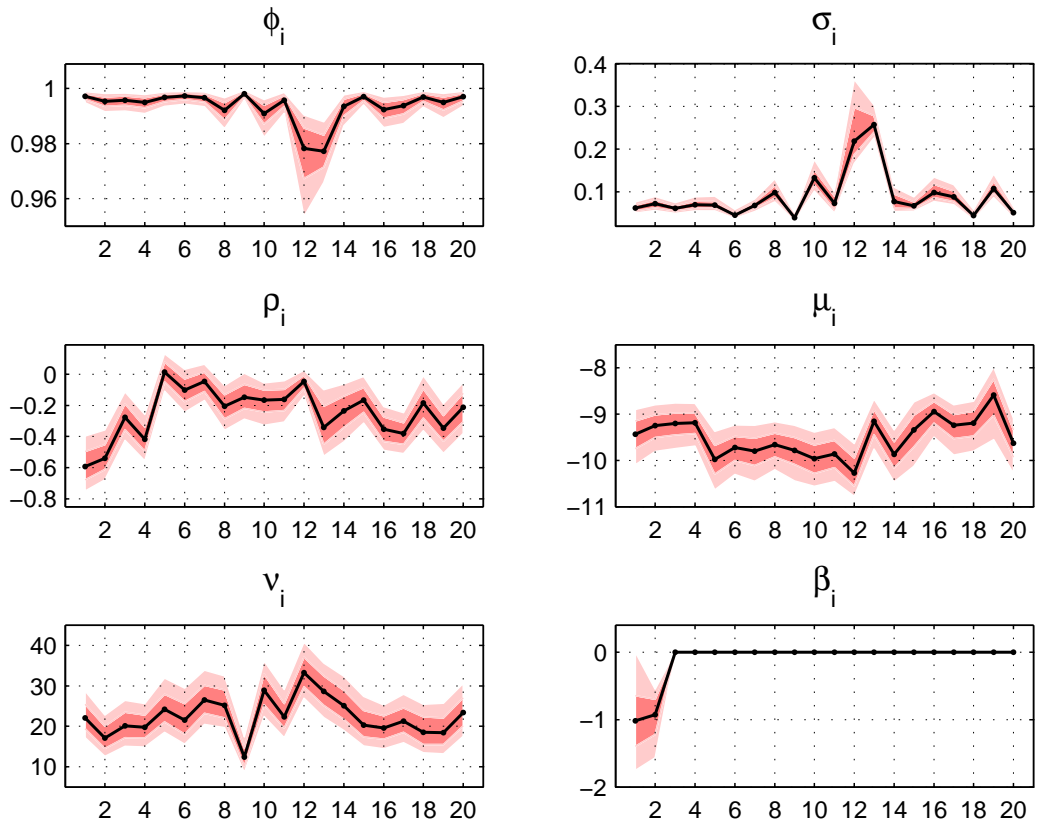} 

\caption{Posterior estimates for parameters from Model CSS: Posterior medians (solid line) and 50\% (filled area, dark) and 80\% (light) credible intervals. The horizontal axis refers to the series index $i$.}

\label{fig-parw}
\end{figure}

\begin{figure}[p]

\centering
\includegraphics[width=5.5in]{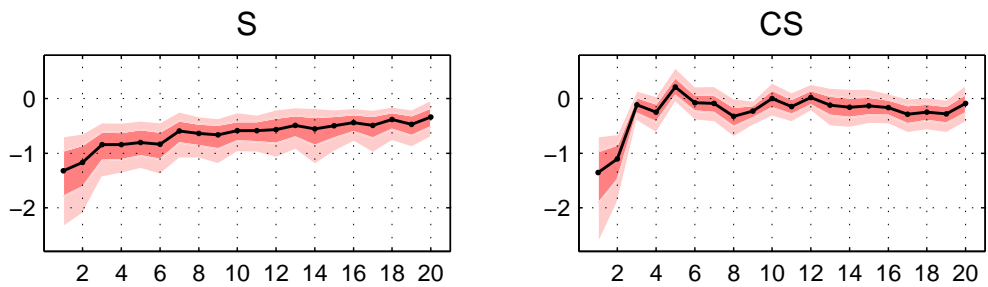}

\caption{Posterior estimates for $\beta_i$: Posterior medians (solid line) and 50\% (filled area, dark) and 80\% (light) credible intervals. The horizontal axis refers to the series index $i$.}

\label{fig-betaw}
\end{figure}

Figure \ref{fig-parw} shows posteriors of the model parameters for Model CSS. A remarkable evidence is considerable shrinkage of $\beta_i$'s except for the first two series, suggesting much parsimonious structure induced by the skew selection. Figure \ref{fig-betaw} plots posteriors of $\beta_i$ for Models S and CS. Note that the series are ordered by posterior means of $\beta_i$'s based on Model S, although the posterior medians displayed here are not monotonically increasing. Model CS exhibits interesting estimates; a posterior distribution of $\beta_5$ leans to positive, presumably for adjusting the skewness in connection with the former series on the Cholesky-type compound processes. This finding and the evidence of shrinkage in Model CSS suggest that the skewness parameter can be redundant in the correlated MSV models.

Other parameters in Figure \ref{fig-parw} show some differences in behaviors of stock price indices among the countries. The series of Spain ($i=12$) and Russia ($i=13$) exhibit smaller $\phi_i$'s and larger $\sigma_i$'s, implying less persistent volatility dynamics. The series of EUR ($i=1$) and US ($i=2$) show large leverage effects with posterior medians of $\rho_i$ below $-0.5$. An important advantage here is that the proposed Cholesky MSV models easily scales up in its dimension with the reduction of posterior computation to univariate stochastic volatility analysis.


\newpage

\section{Concluding remarks}

A new framework of building correlated multivariate stochastic volatility models with skew distributions is developed. The approach of Cholesky-type covariance structure effectively embeds the univariate stochastic volatility with leverage effects and the GH skew $t$-distributions to the multivariate analysis. The salient feature of the proposed model is the skew selection based on the sparsity prior on the skewness parameters. In stock return analyses, the empirical evidence shows the sparse skew and dynamic correlated structures contribute to improved prediction ability in terms of the predictive density and portfolio VaR forecasts, which is practically relevant to business and policy uses of such models in investment and risk management.

There are a number of methodological and computational areas for further investigation. In terms of modeling strategy, the sparse skew structure can be applied to factor stochastic volatility models, which have been widely studied in literature~\citep{GewekeZhou96,PittShephard99b,AguilarWest00,ChibNardariShephard06}. Also, the time-varying sparsity technique using latent threshold models proposed by \cite{NakajimaWest12jbes,NakajimaWest12jfe} can be employed to explore more parsimonious covariance structure for the skew MSV models. One important open question is a potential computational strategy of sequential particle learning algorithms~\citep{Carvalho10} for the proposed MSV models, which would be useful in real-time decision making context.


\medskip

\section*{Acknowledgements}

The author thanks Kaoru Irie for helpful discussion. Computations were implemented using Ox~\citep{Doornik06}.


\medskip

\section*{Appendix. Prior sensitivity analysis}

A forecasting study with different prior distributions is examined for the S\&P500 Sector Indices data used in Section \ref{sec:sp}. Consider the following priors: (Prior-1) $\kappa \sim B(2, 8)$, (Prior-2) $\beta_i \sim \kappa N(\beta_i|-1, 2) + (1-\kappa)\delta_0(\beta_i)$, and (Prior-3) $\nu_i \sim G(24, 0.6)I[\nu_i>4]$. All the other parameters remain the same as the baseline priors specified in Section \ref{sec:data}. Compared to the baseline priors, the new priors imply more concentrated densities of $\kappa$ (Prior-1) with posterior mean $0.2$, shifted from $0.5$ in the baseline, $\beta_i$ (Prior-2) with posterior mean $-1$ from $0$, and $\nu_i$ (Prior-3) with posterior mean $40$ from $20$.

\begin{table}[t]
\centering
\begin{tabular}{llrrrrrr}

\hline
      &		&	\multicolumn{5}{c}{Horizon ($d$ days)}	&	\\
Model & Prior	&	 1     &  2     &  3     &  4	 &  5      & Total \\
\hline
C     &	(1)	&  34.0   & 54.7   &  45.5   & 52.8  &  50.9   & 237.9 \\
      &	(2)	&  34.6   & 54.0   &  44.1   & 50.9  &  50.2   & 233.8 \\
      &	(3)	&  48.6   & 63.1   &  58.3   & 63.6  &  68.1   & 301.7 \\
CS    &	(1)	&  88.5   & 217.6  &  84.7   & 240.6 &  143.9  & 775.2 \\
      &	(2)	&  89.9   & 223.5  &  81.5   & 241.5 &  148.1  & 784.6 \\
      &	(3)	&  101.6  & 222.7  &  97.0   & 250.9 &  162.6  & 834.7 \\
CSS   &	(1)	&  96.8   & 232.1  &  94.3   & 256.8 &  158.6  & 838.6 \\
      &	(2)	&  95.7   & 227.1  &  89.2   & 246.5 &  150.0  & 808.4 \\
      &	(3)	&  117.2  & 245.6  &  116.3  & 276.2 &  181.6  & 937.0 \\

\hline

\end{tabular}

\caption{Prior sensitivity analysis: Cumulative log predictive density ratios $\mathrm{LPDR}_t(d)$ relative to Model S.}
\label{tab-dens-pri}
\end{table}

Table \ref{tab-dens-pri} reports the cumulative log predictive density ratios of Models CS and CSS relative to Model S, which shows little difference among the different priors. In addition, the VaR forecasts are also computed in the same way as in Section \ref{sec:forecast}; the likelihood ratio tests indicate the null hypothesis that the expected ratio of violations is equal to $\alpha$ is not rejected for Models CS and CSS with the those priors in any case considered in Table \ref{tab-test} at the 5\% significance level. These findings indicate that the results of forecasting performance improved by the skew and correlated structure in the MSV models are quite robust regardless of prior specifications for those key hyper-parameters. \\


\bibliographystyle{asa}
\bibliography{svskm}

\end{document}